\documentclass[12pt]{iopart}
\usepackage{iopams} 
\usepackage{graphics}
\usepackage{pennames}
\usepackage{amssymb}
\usepackage{setstack}
\begin{document}
\title{
{\bf Multiplicity of charged particles in Pb--Pb collisions at SPS energies}
}
%
%

\author{  
The NA57 Collaboration:\\  
F~Antinori$^{l}$, 
P~Bacon$^{e}$, 
A~Badal{\`a}$^{g}$, 
R~Barbera$^{g}$,
A~Belogianni$^{a}$, 
A~Bhasin$^{e}$, 
I~J~Blood\-worth$^{e}$, 
M~Bombara$^{i}$, 
G~E~Bruno$^{b}$,
S~A~Bull$^{e}$,
R~Caliandro$^{b}$,
M~Campbell$^{h}$,
W~Carena$^{h}$,
N~Carrer$^{h}$,
R~F~Clarke$^{e}$,
A~Dainese$^{l}$,
A~P~de~Haas$^{s}$,
P~C~de~Rijke$^{s}$,
D~Di~Bari$^{b}$,
S~Di~Liberto$^{o}$,
R~Divi\`a$^{h}$,
D~Elia$^{b}$,
D~Evans$^{e}$,
G~A~Feofilov$^{q}$,
R~A~Fini$^{b}$,
P~Ganoti$^{a}$,
B~Ghidini$^{b}$,
G~Grella$^{p}$,
H~Helstrup$^{d}$,
K~F~Hetland$^{d}$,
A~K~Holme$^{k}$,
A~Jacholkowski$^{g}$,
G~T~Jones$^{e}$,
P~Jovanovic$^{e}$,
A~Jusko$^{e}$,
R~Kamermans$^{s}$,
J~B~Kinson$^{e}$,
K~Knudson$^{h}$,
A~A~Kolozhvari$^{q}$,
V~Kondratiev$^{q}$,
I~Kr\'alik$^{i}$,
A~Krav\v c\'akov\'a$^{j}$,
P~Kuijer$^{s}$,
V~Lenti$^{b}$,
R~Lietava$^{e}$,
G~L\o vh\o iden$^{k}$,
V~Manzari$^{b}$,
G~Martinsk\'a$^{j}$,
M~A~Mazzoni$^{o}$,
F~Meddi$^{o}$,
A~Michalon$^{r}$,
M~Morando$^{l}$,
E~Nappi$^{b}$,
F~Navach$^{b}$,
P~I~Norman$^{e}$,
A~Palmeri$^{g}$,
G~S~Pappalardo$^{g}$,
B~Pastir\v c\'ak$^{i}$,
J~Pi\v s\'ut$^{f}$,
N~Pi\v s\'utov\'a$^{f}$,
R~J~Platt$^{e}$, 
F~Posa$^{b}$,
E~Quercigh$^{l}$,
F~Riggi$^{g}$,
D~R\"ohrich$^{c}$,
G~Romano$^{p}$,
K~\v{S}afa\v{r}\'{\i}k$^{h}$,
L~\v S\'andor$^{i}$,
E~Schillings$^{s}$,
G~Segato$^{l}$,
M~Sen\'e$^{m}$,
R~Sen\'e$^{m}$,
W~Snoeys$^{h}$,
F~Soramel$^{l}$
\footnote {Permanent address: University of Udine, Udine, Italy},
M~Spyropoulou-Stassinaki$^{a}$,
P~Staroba$^{n}$,
T~A~Toulina$^{q}$,
R~Turrisi$^{l}$,
T~S~Tveter$^{k}$,
J~Urb\'{a}n$^{j}$,
F~F~Valiev$^{q}$,
A~van~den~Brink$^{s}$,
P~van~de~Ven$^{s}$,
P~Vande~Vyvre$^{h}$,
N~van~Eijndhoven$^{s}$,
J~van~Hunen$^{h}$,
A~Vascotto$^{h}$,
T~Vik$^{k}$,
O~Villalobos~Baillie$^{e}$,
L~Vinogradov$^{q}$,
T~Virgili$^{p,}$\footnote{Corresponding author: tiziano.virgili@sa.infn.it},
M~F~Votruba$^{e}$,
J~Vrl\'{a}kov\'{a}$^{j}$\ and
P~Z\'{a}vada$^{n}$.
}

\vspace{3.0cm}

{\small
$^{a}$ Physics Department, University of Athens, Athens, Greece\\
$^{b}$ Dipartimento IA di Fisica dell'Universit{\`a}
       e del Politecnico di Bari and INFN, Bari, Italy \\
$^{c}$ Fysisk Institutt, Universitetet i Bergen, Bergen, Norway\\
$^{d}$ H{\o}gskolen i Bergen, Bergen, Norway\\
$^{e}$ University of Birmingham, Birmingham, UK\\
$^{f}$ Comenius University, Bratislava, Slovakia\\
$^{g}$ University of Catania and INFN, Catania, Italy\\
$^{h}$ CERN, European Laboratory for Particle Physics, Geneva, Switzerland\\
$^{i}$ Institute of Experimental Physics, Slovak Academy of Science,
       Ko\v{s}ice, Slovakia\\
$^{j}$ P.J. \v{S}af\'{a}rik University, Ko\v{s}ice, Slovakia\\
$^{k}$ Fysisk Institutt, Universitetet i Oslo, Oslo, Norway\\
$^{l}$ University of Padua and INFN, Padua, Italy\\
$^{m}$ Coll\`ege de France, Paris, France\\
$^{n}$ Institute of Physics, Prague, Czech Republic\\
$^{o}$ University ``La Sapienza'' and INFN, Rome, Italy\\
$^{p}$ Dipartimento di Scienze Fisiche ``E.R. Caianiello''
       dell'Universit{\`a} and INFN, Salerno, Italy\\
$^{q}$ State University of St. Petersburg, St. Petersburg, Russia\\
$^{r}$ IReS/ULP, Strasbourg, France\\
$^{s}$ Utrecht University and NIKHEF, Utrecht, The Netherlands
%
%
}
\vspace{3.0cm}

\begin{abstract}
The multiplicity of charged particles in the central rapidity region
has been measured by the NA57 experiment in Pb--Pb collisions at the CERN SPS 
at two beam momenta: 158 A GeV/{\it c} and 40 A GeV/{\it c}. 
The value of $dN_{ch}/d\eta$ at the maximum has been determined and its
behaviour as a function of centrality has been studied in the centrality
range covered by NA57 (about 50\% of the inelastic cross section). 
The multiplicity increases approximately logarithmically with the centre 
of mass energy. 
\end{abstract}

\newpage
\section{Introduction}
Ultrarelativistic heavy-ion collisions probe matter at extreme values
of temperature 
and energy density, where the phase transition to the Quark
Gluon Plasma phase is expected to occur~\cite{QM02}.
The multiplicity of charged particles produced at central
rapidity is in this context an important global observable, related
to the entropy of the system formed in the collision~\cite{Letessier}
and to the initial energy density~\cite{Bjorken}.

The NA57 experiment measures the event multiplicity with a dedicated 
Microstrip Silicon Detector (MSD).
This measurement is used for the off-line estimate of the 
centrality of the interactions, i.e. the number of nucleons that participate
in the collision~\cite{Nicola}. This allows us to study the production of
hyperons ($\Lambda$, $\Xi$, $\Omega$ and their antiparticles) in Pb--Pb
collisions as a function of the collision centrality~\cite{Ladislav}. 

In this paper we use the charged particle multiplicity measured by the
MSD to determine the maximum of the pseudorapidity distribution 
($dN_{ch}/d\eta|_{max}$); in the following we describe the procedure employed 
to extract the peak value of $dN_{ch}/d\eta$ from the raw data 
at both beam momenta explored by NA57, 158 $A$ GeV/{\it c} and 40 $A$ 
GeV/{\it c}. The peak value of $dN_{ch}/d\eta$ is the variable most 
frequently used to characterize 
the multiplicity of the interaction, since it allows a comparison between 
different experiments, being independent of the detailed phase space 
acceptance. The measurement of the centrality for the sample at 40 $A$ 
GeV/{\it c}, presented here for the first time, is performed 
with a similar procedure as for the 158 $A$ GeV/{\it c} sample~\cite{Nicola}.
Our measurements cover a centrality range corresponding to the most 
central $53$\% of the total inelastic cross section. 

\section{The experimental set--up}
\label{par2}
The NA57 experiment is designed to study the production of strange particles  
in heavy ion collisions by reconstructing their charged decay products 
in a high granularity silicon pixel telescope placed
in the magnetic field provided by the GOLIATH magnet 
(1.4 T central value).
The complete NA57 set--up has been described elsewhere~\cite{NA57}; here we 
concentrate on the detectors that are relevant for the present analysis. 

NA57 has collected data with a lead beam incident on a lead target (1\% 
interaction length) at two beam momenta: 158 $A$ GeV/{\it c}
and 40 $A$ GeV/{\it c}. 
A centrality trigger provided by six scintillators, placed 
around the beam line, 10 cm downstream of  the target, selects about 60\% 
of the most central total inelastic cross section. A special effort was
made to reduce background sources in order to extend the
centrality range towards peripheral events, the main limit coming from the
empty target contamination. Care was also taken to reject double 
interaction events. 

\begin{figure*}
 \centering
\resizebox{0.95\textwidth}{!}{%
  \includegraphics{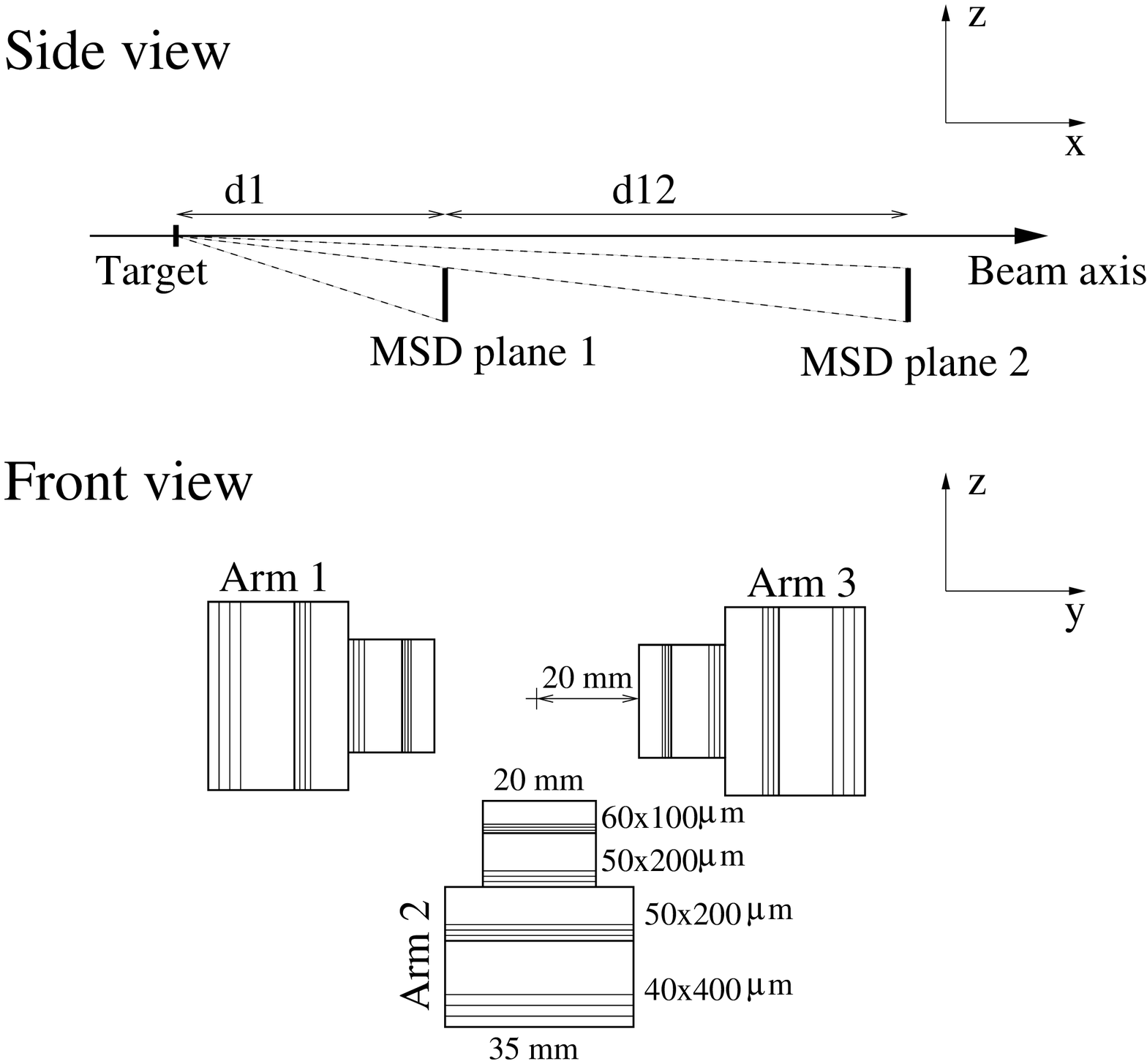}}
 \caption{Layout of the Multiplicity Silicon Detectors (MSD).
   Only the lower arm is shown in the side view. See text for the values
   of the distances {\rm d1} and {\rm d12}.}
 \label{fig:MSDlayout}
\end{figure*}

Two planes of microstrip silicon detectors (MSD) placed  
between the target and the telescope are used to sample the 
charged particle multiplicity. 
Each plane consists of three arms (fig.~\ref{fig:MSDlayout}), 
each of them composed of 200 strips of pitches ranging from 100$\mu$m to 
400$\mu$m. The strip geometries were chosen so as to keep the occupancy 
approximately uniform ($\simeq$ 10\% for the most central Pb-Pb collisions). 
The magnetic field direction is orthogonal to the beam direction and 
parallel to the MSD planes.

The detector efficiency has been determined to be well above 99\% 
by correlating the tracks reconstructed in the telescope 
with the hits recorded in the bottom arm in a special proton beam run where 
the MSD and the telescope were aligned on the beam. 

In the 158 $A$ GeV/{\it c} configuration the two MSD planes are placed 
respectively $19.8$ cm and $54.5$ cm downstream of the target; 
in the 40 $A$ GeV/{\it c} configuration, the planes are positioned at
$20.4$ cm and $38.0$ cm of the target.  
With this geometry, the first and the second plane cover approximately
the pseudorapidity regions $ 1.9 < \eta < 3$ and $3 < \eta < 4$ for the 
158 $A$ GeV/{\it c} configuration (central rapidity $y_{cm}=2.9$) , and $1.9 < \eta < 3$ and 
$2.4 < \eta < 3.6$ for the 40 $A$ GeV/{\it c} configuration (central 
rapidity $y_{cm}=2.2$). 
The transverse momentum acceptance region extends down to a few MeV/{\it c}; 
the azimuthal acceptance is about 30\% for both planes. 

\section{Multiplicity reconstruction}
\label{par3}

The reconstruction proceeds as described in the following. 
The detector provides analogue signals approximately proportional to the 
energy lost by the crossing particle.
A run-by-run calibration\footnote {In normal conditions a run corresponds
to about half an hour of data taking.} 
is performed in order to equalize the strip
signals and to remove the noisy strips (less than 1\% ). 
The number of clusters is determined, where a
cluster is a group of contiguous strips above threshold. 
 
The multiplicity of charged particles on the detector (hit multiplicity,
$N_{hit}$) is then evaluated by an algorithm that takes
into account the total energy deposited in the clusters: 
the cluster charge is compared with the expected values from 
1, 2 and more particles, in order to account for clusters produced by 
the passage of more than 1 particle. 
The hit multiplicity distribution presented in this analysis is obtained 
from a sample of about $10^6$ events covering the full data taking period.

The contribution to the triggered sample from interactions in 
air or in other materials along the beam line ({\em empty target}
contamination) was evaluated using data collected without the target
and then subtracted.

The hit multiplicity is corrected for event uncorrelated background hits,
mostly due to $\delta$-rays, estimated from ``beam trigger'' data. 
This contribution amounts to about 5 hits per event over the full 
MSD for both beam momenta. 

\begin{figure*}
 \centering
\resizebox{0.65\textwidth}{!}{%
  \includegraphics{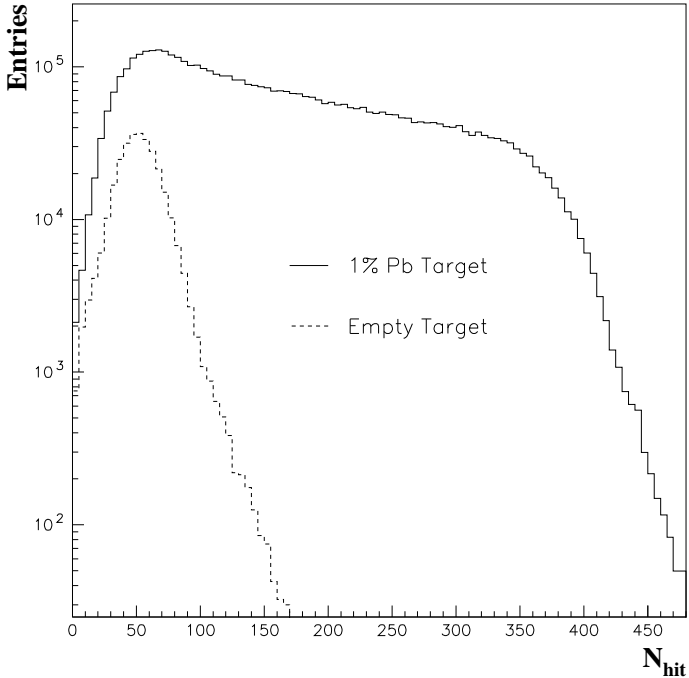}}
\resizebox{0.65\textwidth}{!}{%
  \includegraphics{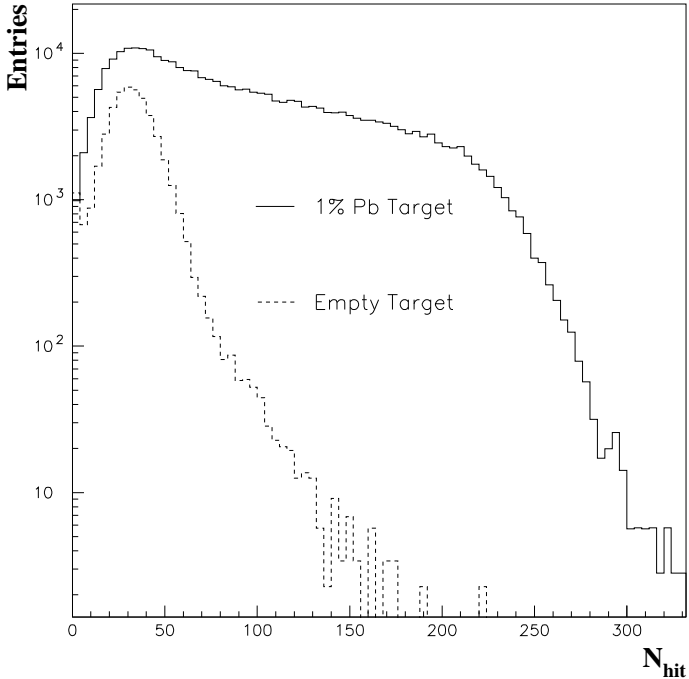}}
 \caption{Hit multiplicity distribution at 158 $A$ GeV/{\it c} (top) 
and 40 $A$ GeV/{\it c} (bottom). The empty target contribution is also shown.}
  \label{fig:158gev}
\end{figure*}

The distribution of the hit multiplicity $N_{hit}$ in the range  
$2 < \eta < 4$ for the 158 $A$ GeV/{\it c} data is shown in 
fig.~\ref{fig:158gev} (top). The decrease at low multiplicity is due to 
the trigger condition. 
The empty target contribution is $\simeq 6\%$, concentrated at very low
multiplicity.

In the 40 $A$ GeV/{\it c} configuration there is an overlap between 
the pseudorapidity coverage of the two planes, corresponding 
to the region $2.4<\eta <3$. In this region the multiplicity is determined
by an average of the multiplicity measured by the two planes. 
A small distortion in the overlap due to the magnetic field is  
corrected for by simulation.
The distribution of $N_{hit}$ in the range $1.9<\eta <3.6$ is shown in 
figure.~\ref{fig:158gev} (bottom).  
The empty target contamination here is $\simeq 13\%$; the $\delta$-ray 
contamination is about the same as at high energy. 
In either case the total measured multiplicity is not 
sensitive to the polarity of the magnetic field. 
 
In order to obtain the charged particle multiplicity produced over the 
full azimuthal range ($N_{\rm ch}$) a geometrical correction must be 
applied. The correction is determined by tracing Pb--Pb events
generated by VENUS~\cite{VENUS} through a GEANT~\cite{GEANT}  
simulation of the apparatus, including the detailed magnetic field
map and the full detector response. 
In this way charge sharing effects, secondary interactions, 
gamma ray conversions and particle decays are accounted for. 
This correction also account for the small extrapolation 
down to $p_t$ equal to zero. 

The resulting distribution of $N_{\rm ch}$ has been used to estimate the 
centrality of the interactions, as discussed in the next section. 

The charged particle multiplicity produced in a restricted range of 
pseudorapidity $\eta$ can be measured by selecting only the strips which 
cover the corresponding acceptance. We indicate by $N_{hit}(|\eta |<0.5)$ 
and $N_{\rm ch}(|\eta |<0.5)$ the hit and the total charged multiplicity  
measured in one unit of $\eta$ at the position of the maximum of the
distribution ($\eta|_{max}$) \footnote {$\eta|_{max}$ =2.4 and 
 $\eta|_{max}$ =3.1 for 40 and 158 $A$ GeV/{\it c} beam momenta 
respectively.}. 

\begin{figure*}
  \centering
\resizebox{0.8\textwidth}{!}{%
  \includegraphics{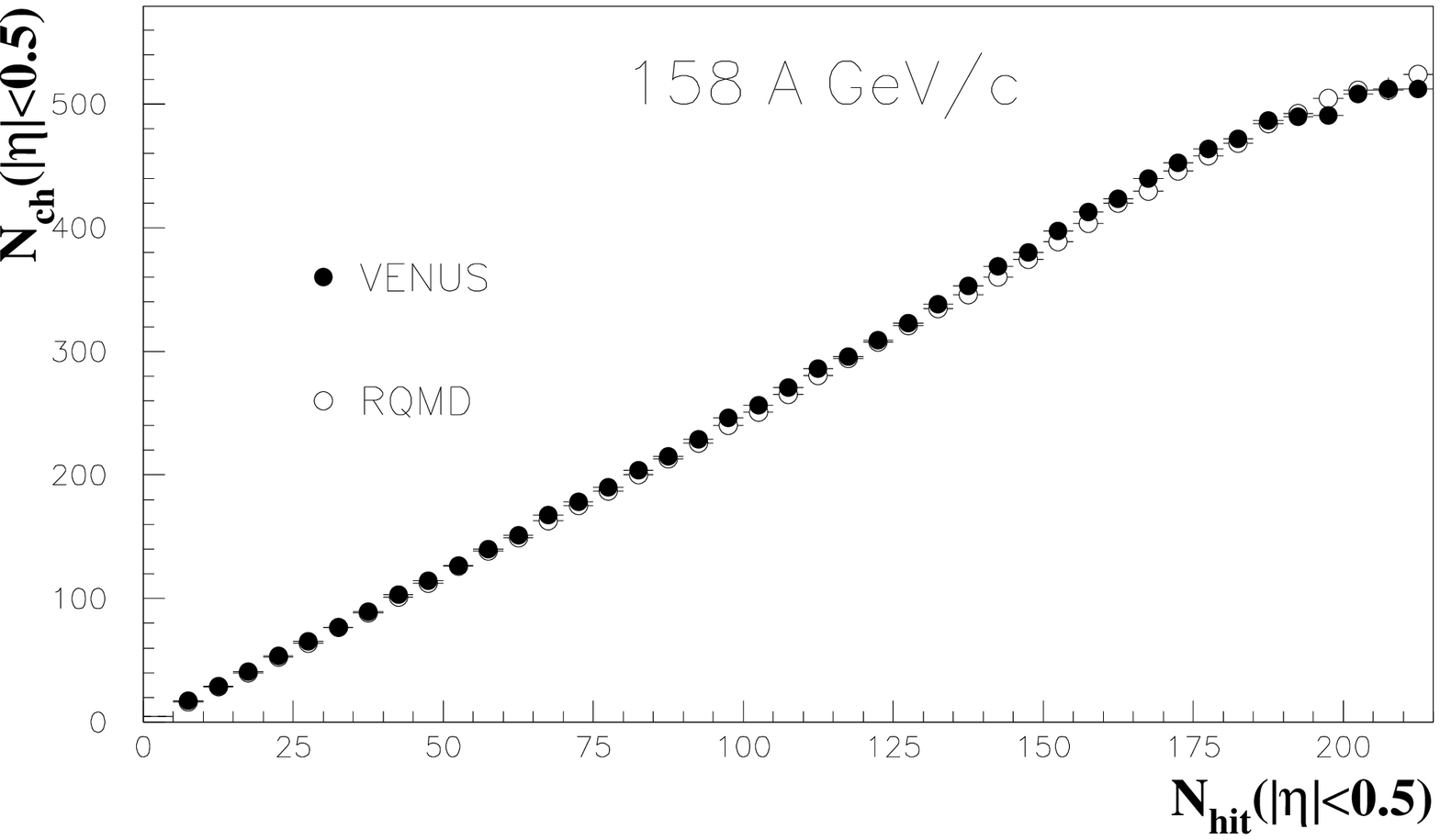}}
\resizebox{0.8\textwidth}{!}{%
    \includegraphics{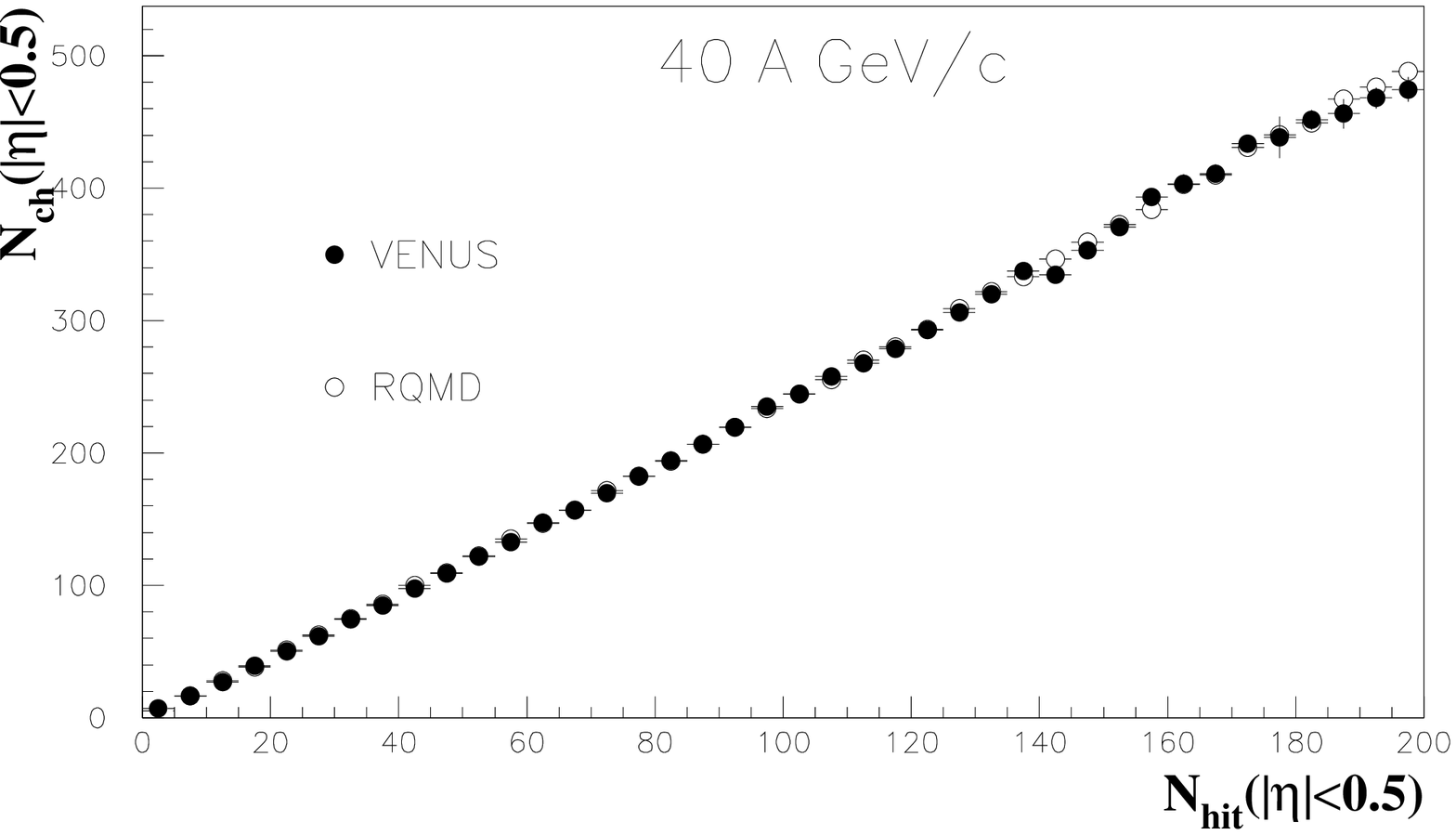}}
\caption{Correlation between charged particle multiplicity and hit 
multiplicity in one unit of pseudorapidity around central rapidity for 
158 $A$ GeV/{\it c} (top) and 40 $A$ GeV/{\it c} (bottom) simulated data.}
  \label{fig:multcorr}
\end{figure*}

In fig.~\ref{fig:multcorr} 
the correlation between $N_{\rm ch}(|\eta |<0.5)$
and $N_{hit}(|\eta |<0.5)$ is shown for the two beam momenta. 
The correlations obtained using two different Monte~Carlo generators 
VENUS and RQMD~\cite{RQMD} are in good agreement; the difference is 
less than 2\% over the full range of multiplicity.

The peak value $dN/d\eta |_{max}$ can be obtained applying a correction
factor $f$ to $N_{\rm ch}(|\eta |<0.5)$, corresponding to the ratio between 
the peak and the average value in the central $\eta$ unit. The factor $f$ 
depends on the shape of the distribution.
Values of the width $\sigma$ of the $\eta$ distribution were measured by 
other collaborations~\cite{NA49}~\cite{NA50}, and a very weak dependence 
on the centrality was reported. This allows us to use the same value 
$f$ for the whole centrality range spanned by NA57. Taking an average of
the reported $\sigma$ values and assuming gaussian shape 
we can estimate $f$=1.02 at 158 $A$ 
GeV/{\it c} and $f$=1.03 at 40 $A$ GeV/{\it c}. 

\section{Multiplicity classes and participants}
\label{par6}
\subsection {158 $A$ GeV/{\it c} data}

\begin{figure*}
 \centering
\resizebox{0.65\textwidth}{!}{%
  \includegraphics{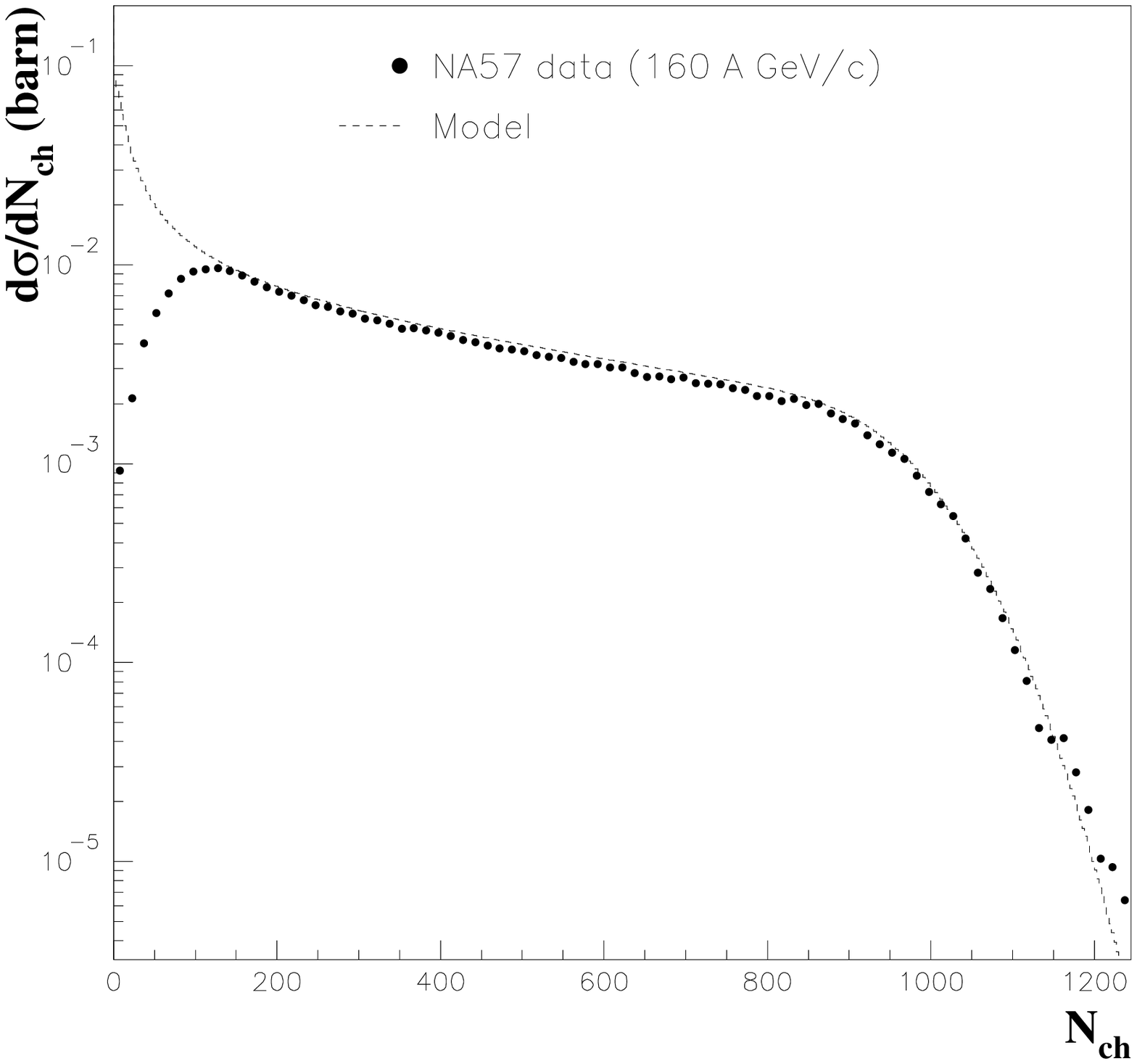}}
\resizebox{0.65\textwidth}{!}{%
    \includegraphics{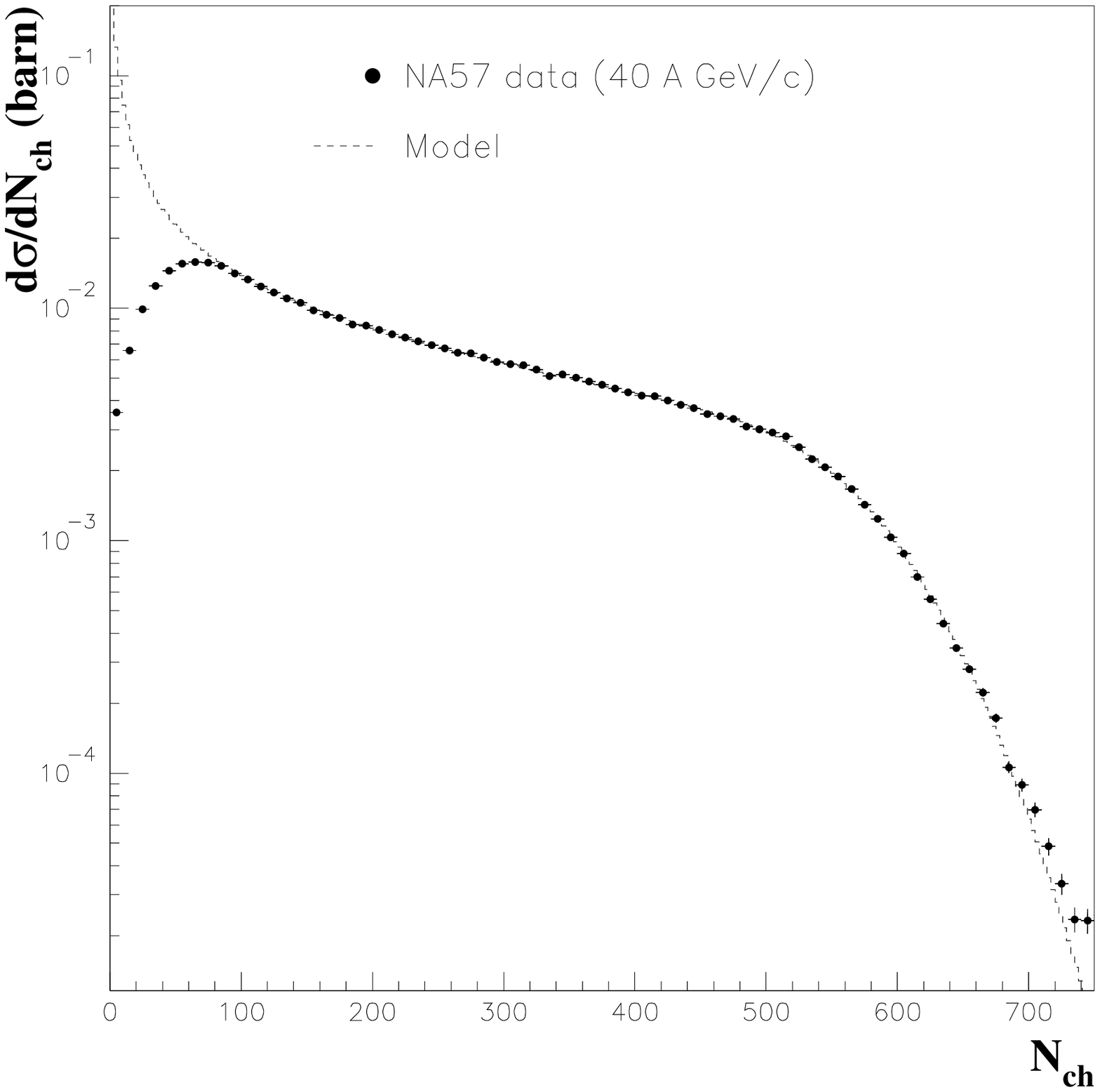}}
\caption{Experimental charged particle multiplicity distributions. Top: 
158 $A$ GeV/{\it c} data (range $2<\eta<4$); Bottom: 
 40 $A$ GeV/{\it c} data (range $1.9<\eta<3.6$). 
The model with $\alpha=1.00$ (top) and  $\alpha=1.10$ (bottom) 
is shown superimposed to the data.}
  \label{fig:figfit158}
\end{figure*}

The distribution of $N_{ch}$, shown in fig.~\ref{fig:figfit158}, is used  
to determine the centrality of the events and to estimate the number of
participants defined as the number of wounded nucleons~\cite{wounded}, 
i.e. the nucleons which suffer at least one primary inelastic collision.  
The procedure to extract the average number of wounded nucleons, 
$<N_{wound}>$, 
has been described in detail in reference~\cite{Nicola}, where  
the multiplicity distribution has been fitted with a model (generalized 
wounded nucleon model), where the charged multiplicity $N_{ch}$ 
is assumed to be proportional to a power of the number of wounded 
nucleons: $N_{ch} \propto N_{wound}^\alpha$. 
The number of wounded nucleons is estimated from a geometrical 
model of the Pb-Pb collision (Glauber model~\cite{wounded}). 

The data are reasonably well described with $\alpha$=1, as assumed in the 
standard 
approach (fig.~\ref{fig:figfit158}, top). Allowing $\alpha$ to vary as a 
free parameter in the fit gives an improved description with values in 
the range 1.02-1.09. This produces a change in $<N_{wound}>$ of 
about one unit; so for the 158 $A$ GeV/c analysis we fix $\alpha$=1.

For the purpose of this study we divide the data into nine centrality 
classes. The centrality is indicated by the corresponding fraction of 
inelastic cross section $\sigma_{inel}$. 
The fractions are evaluated by a MonteCarlo based on the Glauber model, 
and reported in table~1.
The inelastic cross section evaluated by the   
Glauber model, is $\sigma^{glau}_{inel}=7.26$ barn. 
The trigger cross section measured from the rate of triggered 
events, the incoming flux and the target thickness is 
$\sigma^{exp}_{trig}=4.15 \pm 0.11$ barn, in agreement with the 
value obtained with the Glauber fit, therefore the 
trigger cross section is 57\% of the total. 
For this analysis we limit the centrality range to 50\% of $\sigma_{inel}$. 
In table 1, we report for each centrality class, the average 
$dN/d\eta |_{max}$, the average number of wounded nucleons  
$<N_{wound}>$ obtained by the fit (with $\alpha$=1.0) and the r.m.s. of 
their distribution are also reported.

\begin{table}[h]
\caption {Average $dN/d\eta |_{max}$ and $N_{wound}$ as a function of
centrality (158 $A$ GeV/{\it c}).} 
\begin {center}
\begin {tabular} {|c|c|c|c|} \hline
{\bf \% of $\sigma_{inel}$}  & 
{$<dN/d\eta |_{max}>$ } & {$<N_{wound}>$} & {\bf r.m.s.} 
  \\ 
\hline 
0--5   & 478 $\pm$ 10 & 349 $\pm$ 1 & 30 \\
5--10  & 388 $\pm$ 9  & 293 $\pm$ 2 & 33 \\
10--15 & 324 $\pm$ 9  & 245 $\pm$ 3 & 30 \\
15--20 & 266 $\pm$ 8  & 201 $\pm$ 3 & 28 \\
20--25 & 217 $\pm$ 7  & 165 $\pm$ 3 & 25  \\
25--30 & 180 $\pm$ 7  & 136 $\pm$ 4 & 22  \\
30--35 & 149 $\pm$ 6  & 114 $\pm$ 4 & 20  \\
35--45 & 106 $\pm$ 5  &  81 $\pm$ 5 & 20  \\
45--50 &  71 $\pm$ 4  &  51 $\pm$ 5 & 14  \\
\hline 
\end {tabular}
\end {center}
\end{table}

In table 2 the fraction and the absolute values of 
the cross section, integrated over the various centrality ranges are given, 
together with the corresponding average values of $dN/d\eta |_{max}$ and 
$<N_{wound}>$.

\begin{table}[h]
\caption {Cross section, average $dN/d\eta |_{max}$ and $N_{wound}$ 
for different centrality ranges (158 $A$ GeV/{\it c}).} 
\begin {center}
\begin {tabular} {|c|c|c|c|} \hline
 {\bf \% of $\sigma_{inel}$} & {\bf $\sigma$ (barn)}  & 
{$<dN/d\eta |_{max}>$ } & {$<N_{wound}>$} 
  \\ 
\hline 
0--5  & 0.36 $\pm$ 0.01 & 478 $\pm$ 10  & 349  \\
0--10 & 0.73 $\pm$ 0.02 & 434 $\pm$ 10  & 320  \\
0--15 & 1.08 $\pm$ 0.03 & 398 $\pm$ 9   & 295  \\
0--20 & 1.45 $\pm$ 0.03 & 361 $\pm$ 9  & 268 \\
0--25 & 1.82 $\pm$ 0.05 & 332 $\pm$ 8  & 248  \\
0--30 & 2.19 $\pm$ 0.05 & 309 $\pm$ 8  & 231  \\
0--35 & 2.54 $\pm$ 0.07 & 288 $\pm$ 8  & 216  \\
0--45 & 3.27 $\pm$ 0.09 & 244 $\pm$ 7  & 184 \\
0--50 & 3.64 $\pm$ 0.10 & 226 $\pm$ 7  & 170 \\
\hline 
\end {tabular}
\end {center}
\end{table}


In the standard NA57 analysis of the production of strange 
particles, the centrality dependence is studied using 5 
centrality classes~\cite{Ladislav}. 
The resulting values of $<N_{wound}>$ and the r.m.s. of the distributions  
in these classes are given in table~3 together with the empty target 
contribution, as discussed in section 3.

\begin{table}[h]
\caption {NA57 standard centrality classes (158 $A$ GeV/{\it c}).} 
\begin {center}
\begin {tabular} {|c|c|c|c|c|} \hline
 {\bf bin} & {\bf \% of $\sigma_{inel}$}  & {$<N_{wound}>$} & {\bf r.m.s.} 
& {\bf \% of empty target} \\ 
 \hline 
4 & 0--5   & $349 \pm 1$ & 28 & 0 \\
3 & 5--11  & $290 \pm 2$ & 36 & 0 \\
2 & 11--23 & $209 \pm 3$ & 37 & 0 \\
1 & 23--40 & $121 \pm 4$ & 30 & 0.9 \\
0 & 40--53 & $62 \pm 4$  & 18 & 17 \\
 \hline 
\end {tabular}
\end {center}
\end{table}


\subsection {40 $A$ GeV/{\it c} data}

The distribution of $N_{ch}$ (shown in fig.~\ref{fig:figfit158}, bottom),
has been fitted using the same procedure as described above, 
excluding from the
fit the low multiplicity region ($N_{ch}<$110), where the empty target 
contribution is too large. The value $\alpha$=1 at 40 $A$ GeV/{\it c} gives 
a poor fit; the distribution is well described with values of $\alpha$ in 
the range 1.09 to 1.12. The curve drawn in fig.~\ref{fig:figfit158}
corresponds to $\alpha$=1.10. 

The measured trigger cross section is $\sigma^{exp}_{trig}=4.07 \pm 0.16$ 
barn, corresponding to about 56\% of the inelastic cross 
section. This value is compatible with the result obtained from the 
Glauber fit of the multiplicity distribution. 
As in the case of the 158 $A$ GeV/{\it c} data, 
we divide the data into nine centrality classes.
For each centrality class, we indicate the average $dN/d\eta |_{max}$, 
the average 
number of wounded nucleons $<N_{wound}>$ given by the fit 
(with $\alpha$=1.10) and the r.m.s. of the $<N_{wound}>$ distribution.

\begin{table}[h]
\caption {Average $dN/d\eta |_{max}$ and $N_{wound}$ as a function of
centrality (40 $A$ GeV/{\it c}).} 
\begin {center}
\begin {tabular} {|c|c|c|c|} 
\hline
 {\bf \% of $\sigma_{inel}$}  
& {$<dN/d\eta |_{max}>$ } & {$<N_{wound}>$} & {\bf r.m.s.}  
  \\ 
\hline 
0--5   & 348 $\pm$ 10 & 350 $\pm$ 1 & 27 \\
5--10  & 275 $\pm$  9 & 298 $\pm$ 1 & 31 \\
10--15 & 225 $\pm$  8 & 250 $\pm$ 2 & 29 \\
15--20 & 185 $\pm$  7 & 208 $\pm$ 3 & 27 \\
20--25 & 150 $\pm$  6 & 173 $\pm$ 4 & 24 \\
25--30 & 122 $\pm$  6 & 143 $\pm$ 4 & 22 \\
30--35 &  97 $\pm$  5 & 117 $\pm$ 5 & 20 \\
35--45 &  69 $\pm$  4 &  86 $\pm$ 5 & 19 \\
45--50 &  46 $\pm$  3 &  61 $\pm$ 5 & 15 \\
\hline 
\end {tabular}
\end {center}
\end{table}


Table 5 gives the fraction and the absolute values of 
the cross section integrated over various centrality ranges, 
the average values of $dN/d\eta |_{max}$ and $<N_{wound}>$.


\begin{table}[h]
\caption {Cross section, average $dN/d\eta |_{max}$ and 
$N_{wound}$ for different centrality ranges (40 $A$ GeV/{\it c}).} 
\begin {center}
\begin {tabular} {|c|c|c|c|} 
\hline
 {\bf \% of $\sigma_{inel}$}  & {\bf $\sigma$ (barn)}  & 
{$<dN/d\eta |_{max}>$ } & {$<N_{wound}>$} 
  \\ 
 \hline 
0--5  & 0.36 $\pm$ 0.02 & 348 $\pm$ 10 & 346  \\
0--10 & 0.73 $\pm$ 0.03 & 313 $\pm$ 10 & 318  \\
0--15 & 1.09 $\pm$ 0.04 & 283 $\pm$  9 & 295  \\
0--20 & 1.44 $\pm$ 0.05 & 258 $\pm$  9 & 273  \\
0--25 & 1.82 $\pm$ 0.07 & 236 $\pm$  8 & 253  \\
0--30 & 2.18 $\pm$ 0.09 & 217 $\pm$  8 & 235  \\
0--35 & 2.53 $\pm$ 0.11 & 200 $\pm$  7 & 218  \\
0--45 & 3.26 $\pm$ 0.12 & 172 $\pm$  7 & 191  \\
0--50 & 3.64 $\pm$ 0.14 & 159 $\pm$  6 & 177  \\
 \hline 
\end {tabular}
\end {center}
\end{table}

 
The centrality classes used for strange particle analysis at 
40 $A$ GeV/{\it c} are given in table 6. They are defined to give 
the same fractions of total cross section as 
in the 158 $A$ GeV/{\it c} case. 
The $<N_{wound}>$ and their r.m.s. in the case of $\alpha =1.10$ are shown, 
together with the fraction of empty target contamination.

\begin{table}[h]
\caption {NA57 standard centrality classes (40 $A$ GeV/{\it c}).} 
\begin {center}
\begin {tabular} {|c|c|c|c|c|} 
\hline
 {\bf bin} & {\bf \% of $\sigma_{inel}$} & {$<N_{wound}>$} & {\bf r.m.s.} & 
{\bf \% of empty target}
  \\ 
 \hline 
4 & 0--5   & $346 \pm 1$ & 31 & 0 \\
3 & 5--11  & $292 \pm 1$ & 41 & 0 \\
2 & 11--23 & $208 \pm 4$ & 44 & 0 \\
1 & 23--40 & $119 \pm 5$ & 36 & 8.5 \\
0 & 40--53 & $57 \pm 5$ & 24 &  32 \\
 \hline 
\end {tabular}
\end {center}
\end{table}


\subsection {Energy dependence} 

\begin{figure*}
  \centering
\resizebox{0.90\textwidth}{!}{%
  \includegraphics{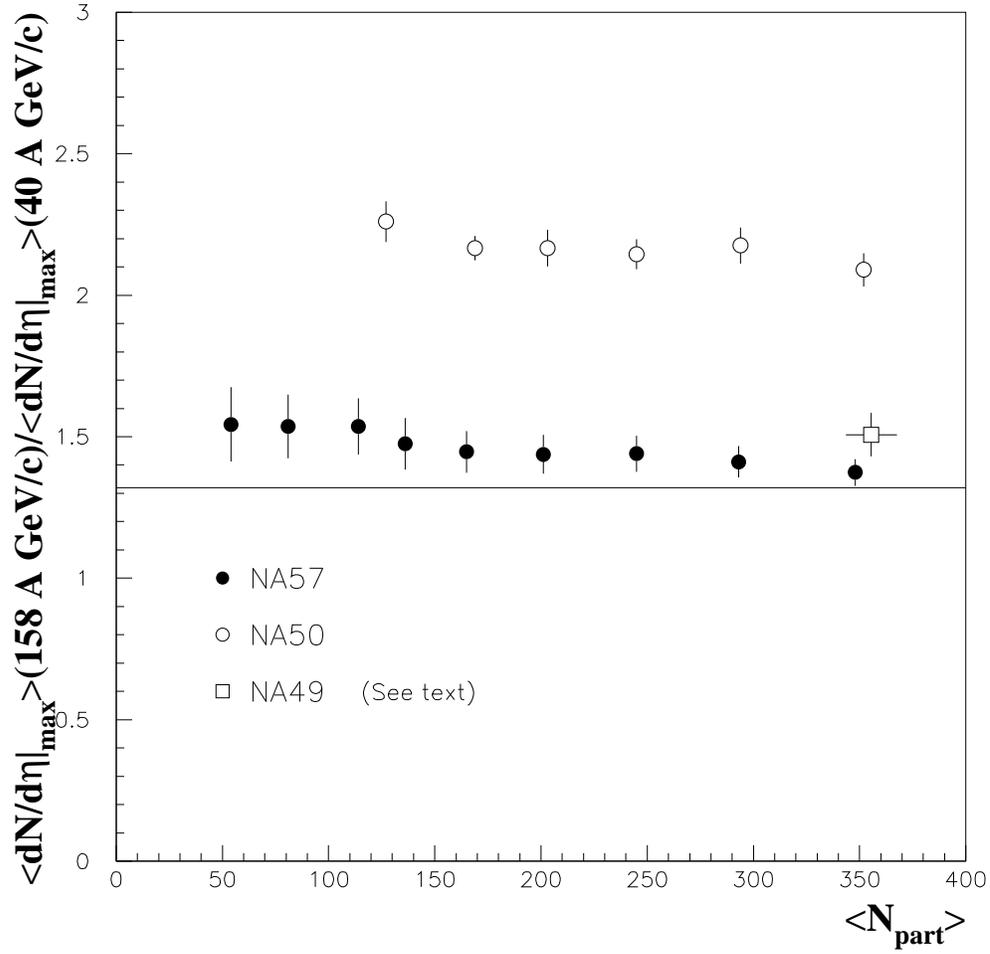}}
\caption{Ratio between $<dN/d\eta |_{max}>$ at 158 and 40 $A$ 
GeV/{\it c} as a function of the participants as measured by 
NA57 (fill circles), NA50 (open circles) and NA49 (open square). 
In the horizontal
scale the average between the participants of the two centrality classes
of NA49 is considered.}
  \label{fig:ratio}
\end{figure*}

In fig.~\ref{fig:ratio} the ratio between $<dN_{ch}/d\eta |_{max}>$ 
at 158 and 40 $A$ GeV/{\it c} beam momenta  
is plotted as a function of $<N_{wound}>$. 
As a result of the above-mentioned difference of the exponent 
$\alpha$ between the two energies, 
the ratio decreases slowly with $<N_{wound}>$, from 1.54 for the most
peripheral to 1.37 for the most central events.  The average ratio is 1.47. 

In proton--proton collisions, the charged multiplicity at 
mid-rapidity is found to scale approximately with the 
logarithm of the centre of mass energy~\cite{Eskola}.  
Assuming the same dependence to hold in Pb-Pb collisions one would expect: 
$dN_{ch}\eta |_{max}$(158 $A$ GeV/{\it c})/$dN_{ch}d\eta |_{max}$ (40 $A$ 
GeV/{\it c})$\simeq \ln (17.3)/\ln (8.77)$=1.31. 
This value is indicated in fig.~\ref{fig:ratio} by a horizontal line. 
In the same figure we also plot data from NA49 and NA50. The comparison 
between different experiments will be discussed in the next session.

\section{Comparison with other experiments}

\begin{figure*}
  \centering
\resizebox{0.90\textwidth}{!}{%
  \includegraphics{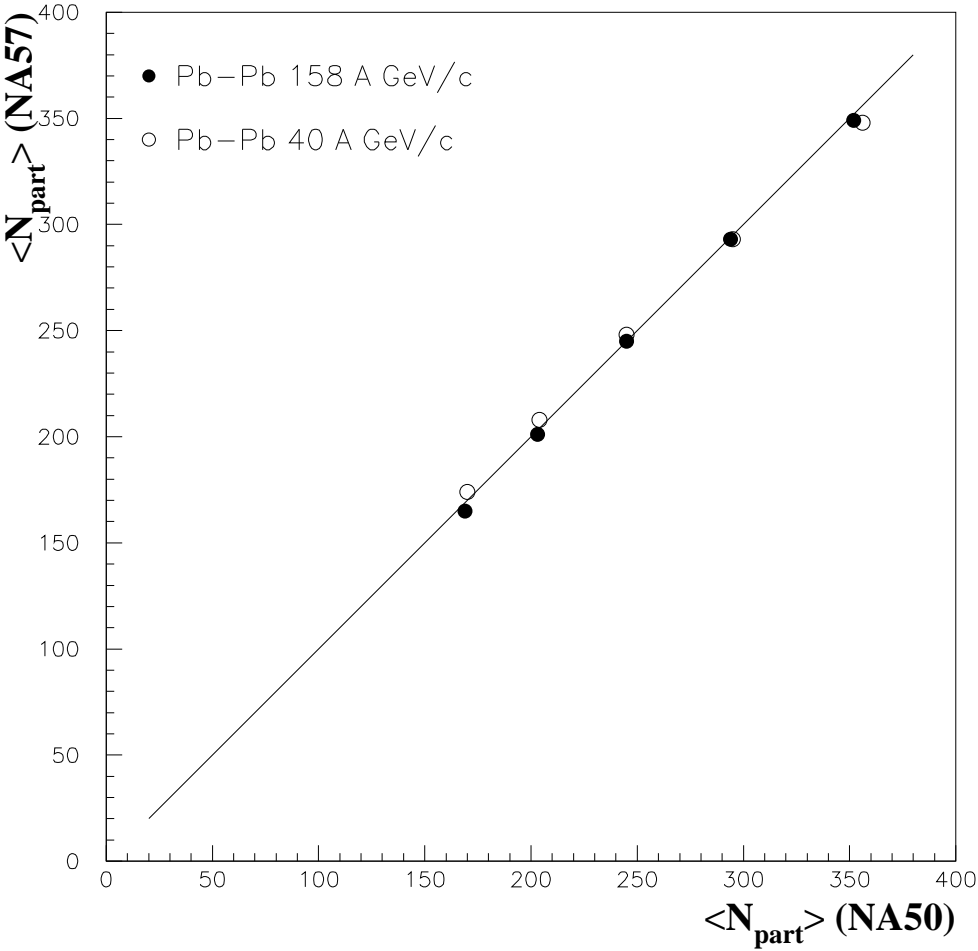}}
\caption{Comparison between the number of participants as determined by 
NA57 and NA50, for both 158 and 40 $A$ GeV/{\it c} data. 
The full line indicates the diagonal; the error bars, 
where not shown, are contained within the symbols}
  \label{fig:part}
\end{figure*}

A comparison between the number of participants as determined by NA57 and 
NA50~\cite{NA50} is shown in fig.~\ref{fig:part}, for both 40  and 
158 $A$ GeV/{\it c} samples. The NA57 points correspond to the first 
five classes of tables 1 and 4, while the NA50 range of centrality is 
35\% of $\sigma_{inel}$. A good agreement is observed.  
A maximum deviation of less than 3\% (of the same order of the measurement 
error) is found in the 20-25\% of $\sigma_{inel}$ class for the 
40 $A$ GeV/{\it c} sample.

\begin{figure*}
  \centering
\resizebox{0.90\textwidth}{!}{%
  \includegraphics{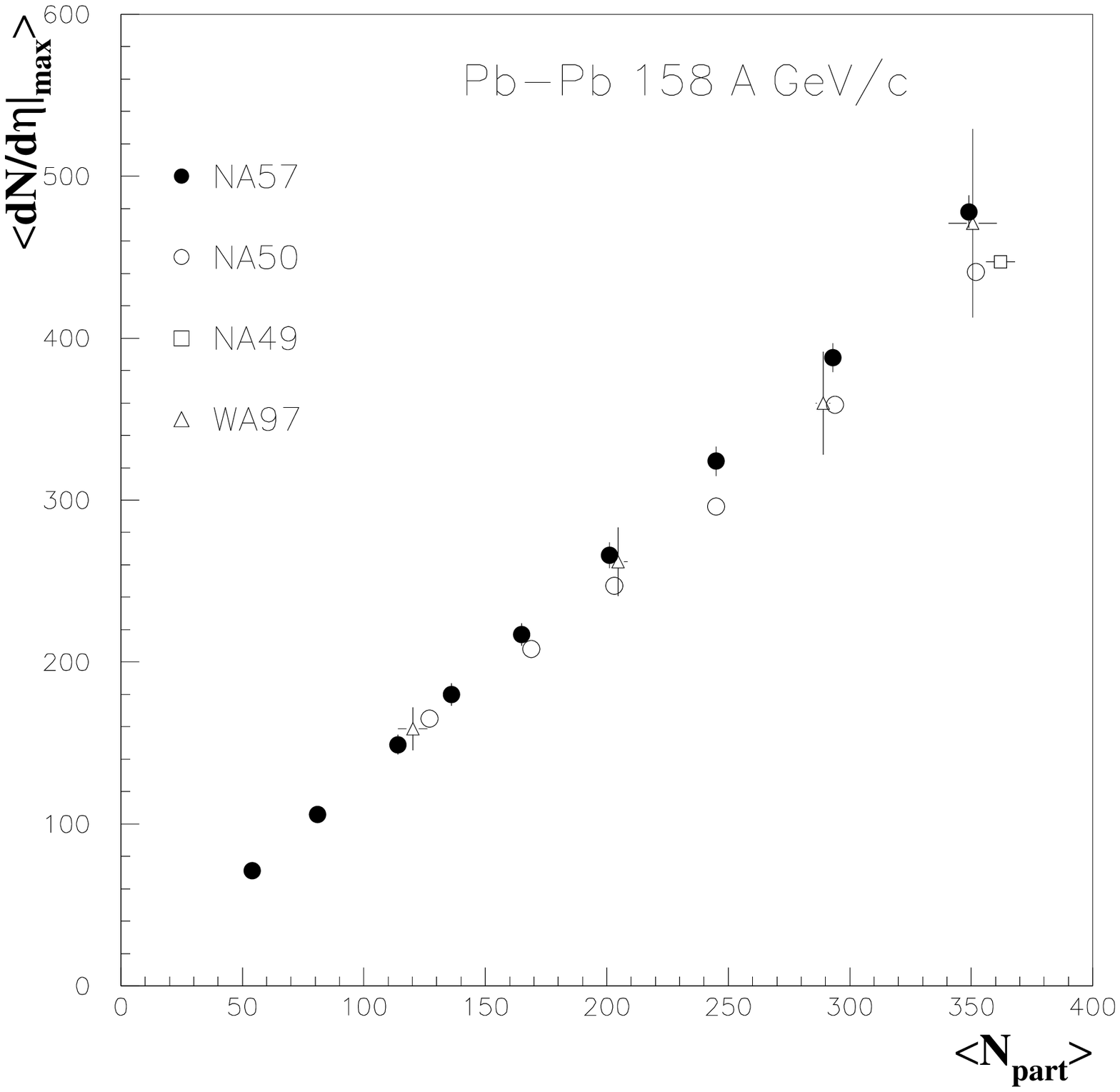}}
\caption{$<dN/d\eta |_{max}>$ as a function of 
the average number of participants for the 158 $A$ GeV/{\it c} data.
The errors bars, where not shown, are contained within the symbols.}
  \label{fig:158gevcomp}
\end{figure*}

In fig.~\ref{fig:158gevcomp} the values of $<dN/d\eta |_{max}>$ from table 1
are compared with those measured by NA50~\cite{NA50} at the same beam 
momentum (158 $A$ GeV/{\it c}).  

There is good agreement for the most peripheral classes, whereas a 
systematic difference appears in the most central 
classes. As a consequence, the slopes of the correlations are slightly 
different. Both are compatible with a linear correlation. 
The value of $\alpha$ found by NA50 is 1.00$\pm$0.01(stat)$\pm$0.04(syst).

The NA49 collaboration has measured the multiplicity of identified 
charged particles at several energies~\cite{NA49}. An estimate to  
the total value of 
$dN/d\eta |_{max}$ can be obtained by summing the contributions from  
$\pi$, $K$, protons and their antiparticles.
For the  158 $A$ GeV/{\it c} data we get $dN/d\eta |_{max}$= 447$\pm$ 3 
for the 5\% most central events (5\% of $\sigma_{inel}$).  
The number of participants reported by the NA49 collaboration (362$\pm$6) 
is larger than what is measured in the corresponding range by NA50 and NA57. 

In order to have a further independent check of the multiplicity measurement 
we compare with the multiplicity of negative particles ($h^-$) 
measured in the WA97 telescope in Pb--Pb interactions at 158 $A$ GeV/{\it c} 
beam momentum. The values found by WA97 were obtained by 
extrapolating the yields measured in the acceptance window of the telescope 
($p_t>300$ MeV/{\it c}) down to $p_t$=0, as described in~\cite{WA97}. 
This procedure does not take into account the effect of the ``low $p_t$ 
enhancement'', and so can underestimate the yields of negatives. 
To account for this, we made a new extrapolation using the 
$p_t$ distribution of the negatives provided by VENUS, 
which is in good agreement with experimental distributions down to 
very low values of $p_t$~\cite{VENUS}. 

In order to scale from the $dN/d\eta_{negatives}$ to the $dN/d\eta_{charged}$ 
we multiply the yield of negatives by the ratio charged/negatives 
$r=2.2$ obtained by NA49~\cite{NA49}. The resulting values are plotted in 
fig.~\ref{fig:158gevcomp} as a function of $<N_{wound}>$. 
A good agreement between NA57 and WA97 is obtained over the whole WA97 
centrality range. 

\begin{figure*}
  \centering
\resizebox{0.90\textwidth}{!}{%
  \includegraphics{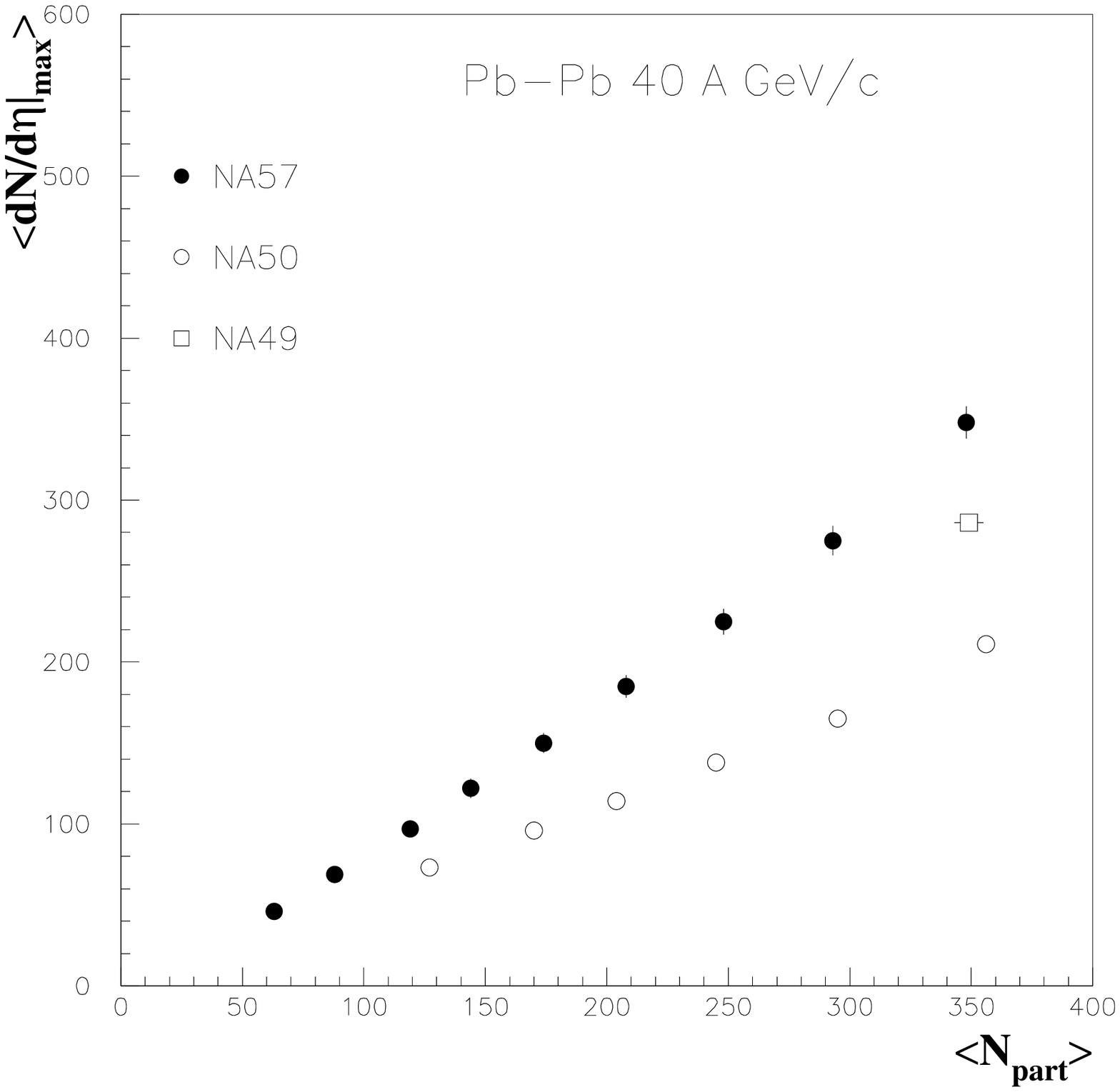}}
\caption{$<dN/d\eta |_{max}>$ as a function of the average 
number of participants for the 40 $A$ GeV/{\it c} data.
The errors bars, where not shown, are contained within the symbols.}
  \label{fig:40gevcomp}
\end{figure*}

Fig.~\ref{fig:40gevcomp} shows the comparison among different experiments 
for the 40 $A$ GeV/{\it c} data. 
We observe that at this energy the slopes of the correlation for NA50 and
NA57 are in strong disagreement. For the most central class there is a 
factor 1.7 between the multiplicity measured by the two experiments. 
As discussed above, we obtain good fits
with values of $\alpha$ in the range $1.09<\alpha<1.12$; the NA50 
collaboration instead reports a value of 
$\alpha$=1.02$\pm$0.02(stat)$\pm$0.06(syst). 

The NA49 collaboration has measured identified charged particles at 
40 GeV~\cite{NA49} for the 7\% most central events (instead of 5\%). 
An estimate of $dN/d\eta |_{max}$ can be determined as before, 
by summing the contributions from  
$\pi$, $K$, protons and their antiparticles. The resulting value, plotted in 
fig.~\ref{fig:40gevcomp}, is 286 $\pm$ 3 particles. 
At 7\% centrality we measure a value of about 320 particles.
Also in this case, the number of participants reported by NA49 
is larger with respect to the determinations made by NA50 and NA57 at the 
same centrality. 

In fig.~\ref{fig:ratio} the ratio between  $<dN_{ch}/d\eta |_{max}>$ 
at 158 and 40 $A$ GeV/{\it c} beam momenta for the NA50 data is shown 
as a function of $<N_{wound}>$. The values are in the range 
2.13-2.38, significantly larger than those measured by NA57. 

The NA49 data at the two energies are given for two different centrality 
ranges. We therefore consider the ratio: 
$${<dN_{ch}/d\eta |_{max}>/<N_{wound}> (158 A GeV/{\it c})}\over  
{<dN_{ch}/d\eta |_{max}>/<N_{wound}> (40 A GeV/{\it c})}$$ 
which gives the value 1.51. 
The horizontal position of the point in
fig.~\ref{fig:ratio} is the average between the participants of the most 
central classes at the two energies. 
As can be seen, the NA49 estimate is close to the measurement made by NA57.

\section{Conclusions}

\label{par7}

We have measured the multiplicity of charged particles  
in Pb--Pb collisions at the CERN SPS at two beam momenta, 158 and 
40 $A$ GeV/{\it c}, in the pseudorapidity 
ranges $2<\eta<4$ and $1.9<\eta<3.6$ respectively. The 
multiplicity distributions are well described by a generalized wounded 
nucleon model where $<N_{ch}>\propto N_{wound}^\alpha$. The 
value of $\alpha$ is compatible with 1 for the 158 $A$ GeV/{\it c} data, and 
slightly larger than 1 for the 40 $A$ GeV/{\it c} data. 
  
The value of $dN_{ch}/d\eta$ at the maximum has been studied 
as a function of centrality over most of the centrality
range covered by NA57.
At 158 $A$ GeV/{\it c} a good agreement is observed with the values 
determined extrapolating the WA97 results on negative particles, 
whereas some discrepancy is observed with NA50 and NA49 for the 
central classes. 
At 40 $A$ GeV/{\it c} a strong disagreement among the three experiments 
is observed, when the measured energy dependence of the charged particle 
multiplicity is considered however, there is a good agreement between 
NA49 and NA57, whereas a disagreement with NA50 still persists.
 
We have compared the values of the participants for a given centrality 
determined by NA50 and NA57 over a wide centrality range, and found 
them to be very similar. 

Finally, we find that the charged multiplicity at central rapidity 
is close to a logarithmic scaling with the centre of mass energy.


\end{document}